\documentstyle[art12,bezier,emlines2]{article}
\normalbaselines
\begin{document}
\voffset=-3cm
\begin{center}{\Large{\bf{Holography and two phases of the QCD vacuum}
}}\footnote
{Talk at the Sixth Workshop on Non-Perturbative QCD, 5-9 June 2001,
American University of Paris}\\[3mm]
{B. P. Kosyakov}\\[3mm]
{Russian Federal Nuclear Center -- VNIIEF, Sarov 607190, Russia}
\end{center}

\begin{abstract}
The holographic principle is often (and hastily) attributed to quantum 
gravity and domains of the Planck size.
Meanwhile it can be usefully applied to problems where gravitation effects 
are negligible and domains of less exotic size.
The essence of this principle is that any physical system can be taken 
to be either classical, placed in a D+1-dimensional spacetime, 
or quantum-mechanical, located in its D-dimensional boundary. 
For example, one believes that a hydrogen atom is a typical quantum system 
living in a four-dimensional spacetime, but it can also be conceived as a
classical system living in a five-dimensional embracing spacetime.
The subnuclear realm is more intricate since  the gluon vacuum reveals
two phases, the hadronic and plasma phases. 
They differ in energetics and symmetry.
Moreover, the classical four-dimensional picture is pertinent to the 
behavior of 
constituent quarks while the plasma phase is expected to be grasped by 
standard four-dimensional QCD.
The relation between the holographic standpoint and the symmetry treatment of 
these two phases is outlined.
Exact retarded solutions to the classical SU(N)
four-dimensional Yang-Mills equations with the source composed of several
point-like colored particles is considered. 
Features of these solutions in the large-N limit provide 
insight into the gauge symmetries of two gluon vacua. 
\end{abstract}

\section{Introduction}
By now, QCD is successful in the weak coupling regime, 
in contrast to the situation in the strong coupling regime. 
To restate this in a different way, one points to the happiness 
in high energy region and troubles with low energy region. 
Despite many efforts, no one managed to continue analytically between 
the weak and strong couplings, or between the ultraviolet and infrared 
energy regions. 
It is reasonable to anticipate a singularity preventing such a 
continuation. 
If this singularity is of the branch point type,
a two-sheeted picture of the subnuclear world emerges. 
So, we deal with two quite different realms, 
{\it cold} and {\it hot}, or, C- and H-realms for short. 
The transition from the former to the latter is a phase transition 
in the thermodynamic sense \cite{MeyerOrtmanns}. 
It now goes under the name {\it deconfinement}. 
The critical energy is estimated to be $E_c\approx 200$ MeV.

According to the general theory  of phase transitions, two adjacent 
phases reveal different symmetries. 
When on the subject of a quantum-field system, the well-known 
theorem by Coleman \cite{Coleman} reads: The invariance of the system is 
the invariance of its ground state, the vacuum. 
Since vacuum averages behave classically, one should examine the 
symmetry of solutions to the field equations in the classical limit. 
In other words, the invariance of the system is the invariance of its 
classical background. 
The vacuum is filled with background fields.
They form a relief relative to which quantum excitations are generated. 
We will see that the gauge symmetry of the H-realm is the usual 
QCD color-$SU(3)$ while that of the C-realm is $SL(4,R)$, 
the Ne'eman--{\v S}ija{\v c}ki symmetry \cite{NeeSij8, NeeSij3}.  

These gauge symmetries should explain essential features of both phases. 
The central idea advocated here is as follows. 
A certain four-dimensional {\it classical} picture is attributed to
the C-realm while the H-realm is described by a four-dimensional 
{\it quantum} picture, the usual QCD picture, which is in the 
holographic correspondence with some five-dimensional classical picture. 
So, geometrically, deconfinement is a topological transition
augmenting the spacetime dimension by one. 
Furthermore, the melting of the cold phase is a quite ordinary event. 
It occurs whenever the acceleration of any cold quark exceeds the critical 
value $a_c$ which is of order of the critical energy $E_c$. 
We will see that deconfinement is indeed associated with a singular point
on the coupling scale and a branch point on the energy scale where the 
spectral condition 
\[p^{2}\geq 0 \]
should have been violated but for other physical sheet, the H-realm sheet. 

\section{Phenomenology}
Let us consider some well known, yet poorly understood facts.
Firstly, the bulk of the hadron phenomenology is grasped by planar diagrams 
\cite{Witten79}.
This implies that world lines of valence quarks are subjected to neither 
bifurcation nor termination (or, more precisely, the world line never
takes a sharp downward turn) unless hadrons collide or decay. 
Such persistence of the quarks is characteristic for classical particles
which are immune from creations and annihilations.
The quark-antiquark sea is largely suppressed in hadrons.
Thus hadrons are systems with a fixed number (two or three) of constituents.
Furthermore, the Okubo--Zweig--Iizuka rule holds that creations and 
annihilations of quark-antiquark pairs inside hadrons are banned 
\cite{Ogawa-et-al}.
For example, the decay $\phi\to K+{\bar K}$ dominates over the 
decay into $\rho+\pi$ which goes through the $s{\bar s}$ annihilation, 
Fig.\ \ref{decay}.
This rule shows that cold quarks differ radically from familiar 
quantum-mechanical objects.
Indeed, if the left diagram in Fig.\ \ref{exchange} is allowable, then the 
right diagram, derivable from it by interchanging the identical quarks,
with due regard for their statistics, should seemingly be also allowed 
but for the Okubo--Zweig--Iizuka rule.
Being sufficiently rigorous, this rule provides evidence that a classical 
picture is pertinent to the behavior of cold quarks.
\begin{figure}[htb]
\begin{minipage}[t]{70mm}
\unitlength=1mm
\special{em:linewidth 0.4pt}
\linethickness{0.4pt}
\begin{picture}(54.00,36.00)
\emline{15.00}{10.00}{1}{15.00}{20.00}{2}
\emline{15.00}{20.00}{3}{12.00}{25.00}{4}
\put(8.00,31.00){\vector(2,-3){4.00}}
\emline{17.00}{10.00}{5}{17.00}{20.00}{6}
\emline{17.00}{20.00}{7}{20.00}{26.00}{8}
\put(20.00,26.00){\vector(1,2){3.00}}
\put(21.00,32.00){\vector(-1,-2){3.00}}
\put(14.00,25.00){\vector(-2,3){4.00}}
\bezier{32}(14.00,25.00)(16.00,22.00)(18.00,26.00)
\put(16.00,6.00){\makebox(0,0)[cc]{$\phi$}}
\put(9.00,35.00){\makebox(0,0)[cc]{$K^{+}$}}
\put(25.00,36.00){\makebox(0,0)[cc]{$K^{-}$}}
\put(13.00,14.00){\makebox(0,0)[cc]{$\bar s$}}
\put(8.00,26.00){\makebox(0,0)[cc]{$\bar s$}}
\put(19.00,14.00){\makebox(0,0)[cc]{$s$}}
\put(24.00,28.00){\makebox(0,0)[cc]{$s$}}
\put(18.00,32.00){\makebox(0,0)[cc]{$\bar u$}}
\put(14.00,29.00){\makebox(0,0)[cc]{$u$}}
\put(45.00,20.00){\vector(0,-1){6.00}}
\put(47.00,10.00){\vector(0,1){5.00}}
\emline{47.00}{15.00}{9}{47.00}{20.00}{10}
\emline{45.00}{14.00}{15}{45.00}{10.00}{16}
\bezier{20}(45.00,20.00)(46.00,22.00)(47.00,20.00)
\put(37.00,31.00){\vector(1,-1){5.00}}
\put(50.00,26.00){\vector(2,3){4.00}}
\bezier{44}(50.00,26.00)(46.00,21.00)(42.00,26.00)
\put(44.00,26.00){\vector(-1,1){5.00}}
\put(52.00,32.00){\vector(-2,-3){4.00}}
\bezier{20}(48.00,26.00)(46.00,24.00)(44.00,26.00)
\put(46.00,6.00){\makebox(0,0)[cc]{$\phi$}}
\put(38.00,35.00){\makebox(0,0)[cc]{$\pi^{-}$}}
\put(56.00,36.00){\makebox(0,0)[cc]{$\rho^{+}$}}
\put(43.00,17.00){\makebox(0,0)[cc]{$\bar s$}}
\put(49.00,17.00){\makebox(0,0)[cc]{$s$}}
\put(38.00,26.00){\makebox(0,0)[cc]{$\bar u$}}
\put(43.00,31.00){\makebox(0,0)[cc]{$d$}}
\put(49.00,33.00){\makebox(0,0)[cc]{$\bar d$}}
\put(54.00,28.00){\makebox(0,0)[cc]{$s$}}
\end{picture}
\caption{$\phi$-meson decay: Allowed (left) and forbidden (right) 
diagrams}
\label
{decay}
\end{minipage}
\hspace{\fill}
\begin{minipage}[t]{70mm}
\unitlength=1mm
\special{em:linewidth 0.4pt}
\linethickness{0.4pt}
\begin{picture}(55.00,32.00)
\emline{10.00}{10.00}{1}{10.00}{13.00}{2}
\put(10.00,17.00){\vector(0,-1){4.00}}
\emline{10.00}{17.00}{3}{7.00}{22.00}{4}
\emline{7.00}{22.00}{5}{7.00}{22.00}{6}
\put(4.00,27.00){\vector(2,-3){3.33}}
\put(12.00,10.00){\vector(0,1){3.00}}
\emline{12.00}{13.00}{7}{12.00}{17.00}{8}
\emline{12.00}{17.00}{9}{12.00}{17.00}{10}
\emline{12.00}{17.00}{11}{15.00}{22.00}{12}
\put(15.00,22.00){\vector(2,3){3.33}}
\put(16.00,28.00){\vector(-2,-3){3.33}}
\put(9.00,23.00){\vector(-2,3){3.33}}
\bezier{36}(9.00,23.00)(10.00,20.00)(13.00,23.00)
\put(7.00,31.00){\makebox(0,0)[cc]{$q_\alpha$}}
\put(22.00,29.00){\makebox(0,0)[cc]{$q_\alpha$}}
\emline{44.00}{10.00}{13}{44.00}{13.00}{14}
\put(44.00,17.00){\vector(0,-1){4.00}}
\emline{44.00}{17.00}{15}{41.00}{22.00}{16}
\put(38.00,27.00){\vector(2,-3){3.33}}
\put(46.00,10.00){\vector(0,1){3.00}}
\emline{46.00}{13.00}{17}{46.00}{18.00}{18}
\put(46.00,18.00){\vector(-2,3){3.33}}
\emline{40.00}{28.00}{19}{43.00}{23.00}{20}
\put(52.00,28.00){\vector(-1,-2){3.00}}
\put(51.00,21.00){\vector(1,2){3.00}}
\bezier{16}(51.00,21.00)(49.00,20.00)(49.00,22.00)
\put(40.00,31.00){\makebox(0,0)[cc]{$q_\alpha$}}
\put(56.00,30.00){\makebox(0,0)[cc]{$q_\alpha$}}
\put(28.00,17.00){\makebox(0,0)[cc]{$\pm$}}
\end{picture}
\caption{Decay where the identical quark exchange is taken into account}
\label
{exchange}
\end{minipage}
\end{figure}

Secondly, the hadronic spectrum can be displayed in the form of 
the Regge trajectories on the Chew--Frautschi plot of 
the mass squared $M^{2}$ versus the angular momentum $J$. 
Hadrons belonging to some Regge trajectory are separated by intervals 
$\Delta J=2$.
Ne'eman and {\v S}ija{\v c}ki related this Regge arrangement to an 
exhaustive group classification of hadrons. 
Regge sequences prove to be associated with infinite multiplets of the 
noncompact group $SL(4,R)$. 
The Lie algebra $sl(4,R)$ is the minimal scheme capable to explain
two features of Regge trajectories: The $\Delta J=2$ rule and the 
infinite sequence of hadronic states.
For more details see \cite{NeeSij8, NeeSij3}.  
To sum up, $SL(4,R)$ seems to have a direct bearing on the gauge invariance 
of the C-realm.

\section{Basic principles 
and specific assumptions}
Since we are interesting in the gluon background fields, we should 
begin with the QCD equations in the limit $\hbar\to 0$.
According to 't Hooft and Witten, the classical limit of QCD is 
related to the limit of large number of colors.
One should substitute $SU(3)$ by $SU(N)$ and go to infinite $N$ to 
arrive at a picture where planar diagrams dominate and vacuum 
fluctuations of gauge invariant operators disappear. 
The exact form of QCD in the large $N$ limit remains in fact unknown. 
I suggest it to be related to the classical $SU(N)$
Yang--Mills--Wong theory as $N\to\infty$ \cite{k8}.

Let us consider classical spinless point particles interacting with the 
classical $SU(N)$ gauge field.
The particles will be called quarks and labelled by index $I$. 
Each quark is assigned a color charge $Q^a_I$. 
Let quarks be moving along timelike world lines $z_I^\mu(\tau_I)$ 
parametrized by the proper times $\tau_I$. 
This gives rise to the current
\begin{equation}
j_\mu (x) =\sum^K_{I=1}\int\!d\tau_I\, Q_I(\tau_I)\,
v^I_\mu (\tau_I)\,\delta^4\biggl(x-z_I (\tau_I)\biggr),
\label{1}\end{equation}                                          % (1)
$Q_I=Q_I^a\,T_a$,\, $T_a$ are generators of $SU(N)$, 
$v^I_\mu\equiv\dot z^I_\mu\equiv dz^I_\mu/d\tau_I$ the 4-velocity.
The action is  
\begin{equation}
S=-\sum_{I=1}^K\,\int\! d\tau_I\,(m^I_0\,\sqrt{v_\mu^I\,v_I^\mu}+
{\rm tr}\,Z_I\lambda^{-1}_I{\dot\lambda}_I)
-\int\! d^4x\,{\rm tr}\,\biggl(j_\mu\,A^\mu+{1\over 16\pi}\,
F_{\mu\nu}\,F^{\mu\nu}\biggr).
\label{2}\end{equation}                                          % (2)
Here, $\lambda_I=\lambda_I(\tau_I)$ are time-dependent elements of $SU(N)$,
$Z_I=e_I^aT_a$,\, $e_I^a$ being some constants whereby the color charge is
specified, $Q_I=\lambda_I Z_I\lambda_I^{-1}$.

The Euler--Lagrange equations are the Yang--Mills equations
\begin{equation}
D^\mu F_{\mu\nu}=4\pi j_\nu,
\label{5}\end{equation}                                           % (5)
the equations of motion of bare quarks
\begin{equation}
m_0^I\,a^\lambda_I=
v_\mu^I\,{\rm tr}\,\biggl(Q_I\, F^{\lambda\mu}(z_I)\biggr),
\label{6}\end{equation}                                           % (6)  
where $a^\lambda_I\equiv \dot v^\lambda_I$ is the 4-acceleration, and the 
Wong equations
\begin{equation}
\dot Q_I=-ig\,[Q_I,\,v^I_\mu\,A^\mu (z_I)]
\label{8}\end{equation}                                           % (8)
describing the evolution of the quark color charges. 
Exact solutions to this set of equations should tell us something
novel about features of the gluon vacuum.

The present approach to the gluon vacuum is much different from the previous
(in particular, the instanton-based one) 
since it takes into account the 
source of the classical Yang--Mills background, 
the classical quarks, which is of crucial 
importance due to the nonlinearity of
the Yang--Mills dynamics. 

\section{Holography}
The holographic principle in its simplest form states:
There exists a correspondence between a given classical picture in a 
${D+1}$-dimensional spacetime and a quantum picture, the $D$-dimensional 
hologram, that occurs in $D$-dimensional 
sections at any fixed instants.
This finding (which went under another slogan) is due to de Alfaro, Fubini and 
Furlan \cite{Alfaro}.
Recall their idea by the example of a system described by 
scalar field $\phi(x)$. 
Let the given quantum system be located in a $D$-dimensional
Euclidean spacetime and specified by a Lagrangian ${\cal L}$. 
One introduces a fictitious time $t$. 
The field becomes a function of the
Euclidean coordinates $x_{1},\ldots ,x_{D}$ and fictitious time $t$, $\phi
=\phi (x,t)$. If ${\scriptstyle\frac{1}{2}}(\partial \phi /\partial t)^{2}$
is treated as the kinetic term, and ${\cal L}$ the potential energy term,
then one defines a new Lagrangian 
\begin{equation}
{\tilde{{\cal L}}}={\scriptstyle\frac{1}{2}}(\partial \phi /\partial t)^{2}-%
{\cal L}
\label
{new-L}
\end{equation}                                           % ({new-L})
generating the evolution in $t$. 
The associated Hamiltonian is 
\begin{equation}
{\tilde{{\cal H}}}={\scriptstyle\frac{1}{2}}\,\pi ^{2}+{\cal L}
\label
{new-H}
\end{equation}                                           % ({new-H})
%where 
$\pi=\partial{\tilde{{\cal L}}}/\partial{\dot{\phi}}=\partial\phi/\partial t$ 
is the conjugate momentum obeying the classical Poisson bracket 
\begin{equation}
\{\phi (x,t),\pi (y,t)\}=\delta ^{D}(x-y).  
\label{P-B}
\end{equation}% (P-B)
It is easy to see that the Gibbs average for an ensemble with the
temperature $kT=\hbar $ 
\begin{equation}
{\cal Z}\lbrack J\rbrack=\int{\cal D}\pi{\cal D}\phi\,\exp\Bigl(-\frac{1}{kT}
\int d^{D}x\,({\tilde{{\cal H}}}+J\phi)\Bigr)  
\label
{Z-cl}
\end{equation}% (Z-cl)
turns to the generating functional for the quantum Green functions 
\begin{equation}
Z\lbrack J\rbrack =\int {\cal D}\phi \,\exp \Bigl(-\frac{1}{\hbar }\int
d^{D}x\,({\cal L}+J\phi )\Bigr)  
\end{equation}% (Z-cl)
upon taking the Gaussian integral over $\pi$. 
The holographic mapping of the bulk picture onto the screen picture 
in the spacetime sections at any instants $t$ is ensured by the Liouville 
theorem: Although $\phi(x,t)$ and $\pi (x,t)$ evolve in $t$, the 
elementary volume in phase space ${\cal D}\pi {\cal D}\phi $ and the Gibbs 
average ${\cal Z}$ are $t$-independent.

Thus, it is meaningless to ask whether a given realm is classical or
quantum. 
It may appear both as classical and quantum, but these two looks
pertain to spacetimes of nearby dimensions. 
To identify the realm, one should only indicate $D$. 
The conventional procedure of quantization only
shifts the seat to another realm: In lieu of the initial classical system
in $D$ dimensions, a new classical system in $D+1$ dimensions emerges. 

The holography provides insight into the origin of quantum anomalies \cite{k0}.
Symmetries of classical Lagrangians may be
sensitive to the dimension; some of them are feasible only for a single 
$D^\star$, while another only for $D=2n$. 
On the other hand, given a
quantized $D^\star$-dimensional theory, we deal actually with the
holographic image of classical $D^\star+1$-dimensional theory, and the
symmetry under examination is missing from it. 

For example, the invariance under the conformal transformations 
$g_{\mu\nu}\to e^{2\varepsilon}\,g_{\mu\nu}$
is known to be attained if the energy-momentum tensor is traceless,
$\Theta_{\hskip1mm\mu}^{\mu}=0$.  
In the classical $D+1$-dimensional Yang-Mills theory, 
\[\Theta^{\mu\nu }\propto(F^{\mu\alpha}F_{\alpha}^{\hskip1.5mm\nu}+
{\scriptstyle\frac{1}{4}}\,\eta^{\mu\nu}F_{\alpha\beta}F^{\alpha\beta}).\]  
The condition $\Theta_{\hskip1mm\mu}^{\mu}=0$ is fulfilled only for $D+1=4$, 
that is Yang-Mills equations are conformal invariant only in a 4D
spacetime.
The quantization of the classical 4D Yang-Mills theory culminates in the
classical 5D Yang-Mills theory where $\Theta^{\mu}_{\hskip1mm\mu}\ne 0$. 
This is how the conformal anomaly and dimensional transmutation 
in QCD occurs.

\section{Solutions to the Yang--Mills--Wong equations}
There are two classes of exact retarded solutions to the 
Yang-Mills equations with point sources \cite{k8}.
Solutions of one class are complex-valued with respect to the Cartan basis 
of $su(N)$, but it is possible to convert them to the real form.
Then such solutions become invariant under $SL(N,R)$ or its subgroups. 
Solutions of another class are real-valued and invariant under $SU(N)$.

The Cartan--Weyl basis of the Lie algebra $su(N)$ is spanned by the 
set of $N^2$ matrices with $N$ elements of the Cartan 
subalgebra, $H_n$, satisfying the relation $\sum_{n=1}^N H_n =0$,
and $N^2-N$ raising and lowering elements $E_{mn}^+$ and $E_{mn}^-$.
The nontrivial commutators are  
\begin{equation}
[H_m,\,E^\pm_{mn}]=\pm\,E^\pm_{mn},\quad [E^+_{mn},\,E^-_{mn}]=H_m -H_n,
\quad [E^\pm_{kl},\,E^\pm_{lm}] =\pm\,E^\pm_{km}.
\end{equation}
One can verify that the Yang--Mills equations are satisfied by 
\begin{equation}
A_\mu=\mp\,{2i\over g}\sum_{I=1}^K\,\biggl(H_I\,{v_\mu^I\over
\rho_I}+g\,\kappa\,E^\pm_{I\,K+1}\,R_\mu^I\,\prod_{I=1}^{K-1}
\delta(R^K\cdot R^I)\biggr).
\label
{A-cold}
\end{equation}
Here, $\kappa$ is an arbitrary real constant, a lightlike vector $R_\mu^I=
x_\mu-z_\mu^I$ is drawn from the 
point of emission of the retarded signal on $I$th worldline, $z_\mu^I$, 
to the point of observation, $x_\mu$, 
$\rho_I=R^I\cdot v^I$ is the retarded distance between these points,
and  $v_\mu^I$ is taken at the retarded instant $\tau_I^{ret}$.
This solution describes a background field generated by 
some $K$-quark cluster in the cold phase.
In addition, there are exact solutions describing the Yang--Mills background 
generated by several quark clusters. 

One can define $(K+1)^2$ traceless imaginary matrices 
${\cal H}_n$ and ${\cal E}_{mn}^\pm$ 
\begin{equation}
{\cal H}_n\equiv i\,H_n,\quad
{\cal E}_{mn}^\pm\equiv i\,E_{mn}^\pm
\end{equation}
which are elements of the Lie algebra $sl(K+1,R)$.
The above solutions become real valued with respect to this basis.
The solutions built from $J^2$ such elements obeying 
the closed set of commutation relations are invariant under 
$SL(J,R)$, $J\le K+1$.
In particular, the Yang--Mills field of a three-quark 
cluster is invariant under $SL(4,R)$, and that of a two-quark 
cluster is invariant under $SL(3,R)$.
Since $SL(3,R)$ is a subgroup of $SL(4,R)$, the background of every 
hadron is specified by the gauge group $SL(4,R)$. 
This symmetry is independent of $N$ and is retained in the limit $N\to\infty$.

For $\kappa =0$, the Yang--Mills equations linearize, and we find a 
Coulomb-like solution corresponding the background field in the hot phase,
\begin{equation}
A_\mu=\sum_{I=1}^K \sum_{n=1}^N\,e_I^n\,H_n\,{v_\mu^I\over\rho_I}
\label
{A-hot}
\end{equation}
with arbitrary $e_I^n$. 
The gauge symmetry of this solution is $SU(N)$.

As is shown in \cite{k9}, in spacetimes of dimension different 
from 4, there is no exact retarded solutions other than Coulomb-like. 
The reason is quite simple.
Only in 4D, both differentiation and multiplication by $A_\mu$
raise the singularity power by one, and hence both terms of the covariant
derivative $\partial_\mu-igA_\mu$ act coherently.

\section{Discussion}
Any background generated by a cold quark occupies individually
some color $sl(2,R)$ cell. 
Neither of two backgrounds generated by different quarks can be 
contained in the same $sl(2,R)$ cell. 
This is similar to the Pauli blocking principle. 
Just as a cell of volume $h^3$ in phase space might be occupied by 
at most one fermion with a definite spin projection, so any color 
$sl(2,R)$ cell is intended for the background of only one quark. 
This color blocking guarantees the totality of color cells against
shrinkage.
We have actually to do with the large-$N$ situation.
We thus arrived at a plausible classical 4D description of the C-realm.
It can be holographically projected onto a 3D quantum picture.

By contrast, the most energetically advantageous field configura\-tion in
the hot phase is such that the color charges of quarks are lined up into a
fixed direction, thereby reducing $SU(N)$ to $SU(2)$. 
We deal with the effectively few-$N$ case. 
The H-realm is described by a 4D perturbative quantum picture
which is holographically dual to a 5D classical picture.

The C-realm and H-realm are linked by a topological transition which
passes through singular points both on the coupling scale [which is seen 
from the comparison of the $g$-dependences of (\ref{A-cold}) and (\ref{A-hot})]
and on the energy scale.

In the C-realm, a dressed quark possesses the 4-momentum \cite{k8}
\begin{equation}
p^{\mu}=m\,(v^\mu+\tau_0\,a^\mu),
\end{equation}
with
\begin{equation}
\tau_0=\frac{2}{3m}\,\vert\,{\rm tr}\,Q^2\,\vert=\frac{8}{3mg^2}\,
(1-\frac{1}{N}).
\label
{tau-0}
\end{equation}
It follows that 
\begin{equation}
p^2=m^2\,\bigl({1+\tau_0^2\,a^2}\bigr).
\end{equation}
The  dressed cold quark should turn to a tachyonic state $p^2< 0$ 
if its acceleration exceeds the critical value 
$\vert a \vert=1/\tau_0$ (which is quite common with hadron collisions).
Such accelerations are 
within the scope of validity of classical description.
Indeed, let two light quarks be separated by distance $\rho$.
The critical acceleration due to the Coulomb-like color force 
is achieved at a distance $\rho\approx \vert {\rm tr}\, Q^2 \vert/m$ 
which, in view of (\ref{tau-0}), is about $g^{-2}$ times greater than 
the Compton wave length of the quarks.
Rather than turn to the tachyonic state, the cold quark  
plunges into the H-realm for the period of the hadron collision. 
The C-realm melts and subsequently the H-realm freezes up again whenever 
the acceleration of any cold quark exceeds $1/\tau_0$, i.\ e., the
energy transferred to it is about its constituent mass $m$, or the 
deconfinement energy $E_c\approx 200$ MeV.

\section{Conclusion and outlooks}
Let me focus your attention on two issues.

Firstly, the {\it classicality} of the C-realm must be verified 
experimentally in the most direct way, for example, with the aid of the 
Bell inequalities.
{The trouble with this project is twofold}.
Technical aspect: it is unclear how to measure kinematical 
characteristics, e.\ g., spin projections,  of individual quarks 
contained in hadrons.
As to the conceptual point of view, the Bell inequalities 
in the current form are independent of the spatial dimension $D$, 
thereby disregarding the
holographic dualism between classical and quantum pictures.
One should render them sensitive to the decision between classical 
and quantum options by just {\it four-dimensional} classical 
observers.

Secondly, the background fields are generated by spinning particles, 
the quarks.
This fact seems to be very important to account for a linkage, 
if it exists, between the deconfinement and the chiral symmetry restoration.
For holographic reasons, there exists a {\it finite} classical 
four-dimensional theory of spinning particles which interact with 
the Yang--Mills field.
It displays such a C-realm where all classical processes are reversible.
\vskip3mm
The financial support of this work by ISTC, project {\#} 840, is acknowledged.

\end{document}